\documentclass[11pt,conference]{IEEEtran}

\ifCLASSINFOpdf
\else
\fi
\usepackage{url}

\usepackage[english]{babel}

\usepackage{xcolor}

\usepackage{soul}

\usepackage[textsize=tiny,colorinlistoftodos]{todonotes}

\usepackage[
        bibstyle=ieee,
        citestyle=numeric-comp,
        maxcitenames=6,
        doi=false,
        eprint=false
    ]{biblatex}
        
\AtEveryBibitem{
  \clearfield{urlyear}%
  \clearlist{language}%
 }

\makeatletter
    \define@key{todonotes}{sm}[]{
        \setkeys{todonotes}{author=\textbf{Sergiy}, color=green!30}}%
\makeatother

\usepackage[range-phrase={\,--\,}, range-units=single, per-mode=symbol, detect-all]{siunitx}

\makeatletter
  \@addtoreset{paragraph}{section}
\makeatother

\usepackage[acronym]{glossaries}
\newacronym{vr}{VR}{Virtual Reality}
\newacronym{ar}{AR}{Augmented Reality}
\newacronym{xr}{XR}{eXtended Reality}
\newacronym{qoe}{QoE}{Quality of Experience}
\newacronym{qos}{QoS}{quality of service}
\newacronym{uav}{UAV}{unmanned aerial vehicle}
\newacronym{faas}{FaaS}{Function-as-a-Service}
\newacronym{pubsub}{Pub/Sub}{Publish/Subscribe}
\newacronym{h3d}{H3D}{holographic 3D}
\newacronym{ai}{AI}{artificial intelligence}
\newacronym{nbmp}{NBMP}{network-based media processing}
\newacronym[plural=KPIs,longplural=key performance indicators]{kpi}{KPI}{key performance indicator}
\newacronym{hc}{\emph{HoloCom}}{\emph{Holographic Communication}}
\newacronym{ml}{ML}{machine learning}
\newacronym{rem}{REM}{radio environment map}
\newacronym{drl}{DRL}{deep reinforcement learning}
\newacronym{fdrl}{FDRL}{federated deep reinforcement learning}
\newacronym{sdr}{SRD}{software-defined radio}
\newacronym[firstplural=radio access technologies]{rat}{RAT}{radio access technology}
\newacronym{vm}{VM}{Virtual Machine}
\newacronym{csi}{CSI}{Channel State Information}
\newacronym{mcs}{MCS}{Modulation and Coding Scheme}
\newacronym{fec}{FEC}{forward error correction}
\newacronym{ue}{UE}{User Equipment}
\newacronym{lstm}{LSTM}{long short term memory}

\usepackage{balance}

\usepackage{cleveref}

\usepackage[all]{nowidow}

\begin{document}
%

\title{True 3D Holography: A Communication Service of Tomorrow and Its Requirements for a New Converged Cloud and Network Architecture on the Path to 6G}



%
\author{
     \IEEEauthorblockN{
        Ingo Friese\IEEEauthorrefmark{1}, Mandy Galkow-Schneider\IEEEauthorrefmark{1},  
        Louay Bassbouss\IEEEauthorrefmark{2}, Alexander Zoubarev\IEEEauthorrefmark{2}, Andy Neparidze\IEEEauthorrefmark{2}, \\
        Sergiy Melnyk\IEEEauthorrefmark{3}, Qiuheng Zhou\IEEEauthorrefmark{3}, Hans D. Schotten\IEEEauthorrefmark{3}\IEEEauthorrefmark{4},
        Tobias Pfandzelter\IEEEauthorrefmark{5}, David Bermbach\IEEEauthorrefmark{5}, \\
        Arndt Kritzner\IEEEauthorrefmark{6}, 
        Enrico Zschau\IEEEauthorrefmark{7}, 
        Prasenjit Dhara\IEEEauthorrefmark{8}, Steve Göring\IEEEauthorrefmark{8}, William Menz\IEEEauthorrefmark{8}, Alexander Raake\IEEEauthorrefmark{8}, \\
        Wolfgang Rüther-Kindel\IEEEauthorrefmark{9}, Fabian Quaeck\IEEEauthorrefmark{9}, Nick Stuckert\IEEEauthorrefmark{9}, Robert Vilter\IEEEauthorrefmark{9},
    }
    \IEEEauthorblockA{
        \IEEEauthorrefmark{1}
        Deutsche Telekom AG, Berlin, Germany \\
        Email: \{ingo.friese, mandy.galkow-schneider\}@telekom.de
    }
    \IEEEauthorblockA{
        \IEEEauthorrefmark{2}
        Fraunhofer FOKUS, Berlin, Germany \\
        Email: \{louay.bassbouss, alexander.zoubarev, andy.neparidze\}@fokus.fraunhofer.de
    }
    \IEEEauthorblockA{
        \IEEEauthorrefmark{3}
        Intelligent Networks,
        German Research Center for Artificial Intelligence (DFKI GmbH), Kaiserslautern, Germany \\
        Email: \{sergiy.melnyk, qiuheng.zhou, hans\_dieter.schotten\}@dfki.de
    }
    \IEEEauthorblockA{
        \IEEEauthorrefmark{4}
        Institute for Wireless Communication and Navigation,
        University of Kaiserslautern-Landau, Kaiserslautern, Germany\\
        Email: schotten@rptu.de
    }
    \IEEEauthorblockA{
        \IEEEauthorrefmark{5}
        Mobile Cloud Computing Group, Technische Universität Berlin \& ECDF, Berlin, Germany \\
        Email: \{tp, db\}@mcc.tu-berlin.de
    }
    \IEEEauthorblockA{
        \IEEEauthorrefmark{6}
        Logic Way GmbH, Schwerin, Germany \\
        Email: kritzner@logicway.de
    }
    \IEEEauthorblockA{
        \IEEEauthorrefmark{7}
        SeeReal Technologies GmbH, Dresden, Germany \\
        Email: ez@seereal.com
    }
    \IEEEauthorblockA{
        \IEEEauthorrefmark{8}
        Audiovisual Technology Group, Technische Universität Ilmenau, Ilmenau, Germany \\
        Email: \{prasenjit.dhara, steve.goering, 
        william.menz, alexander.raake\}@tu-ilmenau.de
    }
    \IEEEauthorblockA{
        \IEEEauthorrefmark{9}
        Aviation Engineering Group, Technical University of Applied Sciences Wildau, Wildau, Germany \\
        Email: \{wkindel, quaeck, stuckert, vilter\}@th-wildau.de
    }
}

\maketitle


\begin{refsection}
    \let\thefootnote\relax\footnote{This is a preprint of a work accepted but not yet published at the 2nd International Conference on 6G Networking (6GNet). Please cite as: \fullcite{friese2023true}.}
\end{refsection}%

\begin{abstract}
Research project \emph{6G NeXt} is considering true 3D holography as a use case, setting requirements on the communication as well as the computing infrastructure. In a future holographic communication service, clients are widely spread in the network and cooperatively interact with each other. Especially for holographic communication high processing power is required as well. This makes a high-speed distributed backbone computing infrastructure, which realizes the concept of split computing, inevitable. Furthermore, tight integration between processing facilities and wireless networks is required in order to provide an immersive user experience. This paper illustrates true 3D holographic communication and its requirements. Afterward, an appropriate solution approach is elaborated. Here, novel technological approaches are discussed based on a proposed overall communication and computing architecture.
\end{abstract}


\section{Introduction}
The \gls{h3d} use case shows a great example of an immersive virtual technology that underlines the need to converge computing and connectivity.
The research project \emph{6G NeXt}~\cite{6gnext} develops an integrated modular architecture enabling future applications to find the most suitable resources in the cloud and network. 
This ability is crucial, especially for an application that needs to split and distribute computing tasks. The proposed architecture involves a set of functionalities, algorithms, and tools for coordinating and managing the execution of these computations across multiple geographically distributed cloud deployments and networks.

The rest of this paper is structured as follows:
Section~\ref{sec:relatedwork} describes existing concepts and how the proposed architecture relates to them. 
Section~\ref{sec:holocom} illustrates Holographic Communication as a use case based on remote rendering as an example of split computing.
Section~\ref{sec:concept} proposes an architecture that supports applications and services to distribute computing tasks across client, edge, and central cloud deployments, taking optimal connectivity into account.
This includes the following parts of the proposed architecture:
\begin{itemize}
    \item[A)] A remote rendering framework as an example of split computing
    \item[B)] A combined high-performance computing and communication backbone helping clients and services to find the most sufficient resources in cloud and network
    \item[C)] Edge-to-cloud services and software extensions to support the architecture
    \item[D)] A \gls{qoe} alignment ensures a great user experience describing the necessary precondition all over the system
    \item[E)] An AI-supported flexible mobile communication platform
\end{itemize}
Section~\ref{sec:conclusion} concludes this paper.

\section{State of the Art and Related Work}
\label{sec:relatedwork}

The idea to split computing tasks of \gls{xr} applications is not new. Many \gls{vr} and \gls{ar} devices do not have sufficient CPU/GPU power and memory to render complex virtual scenes in real time. \emph{3GPP} standardized in release 18 of its 5G specification an \emph{Architecture for enabling Edge Applications}~\cite{3gpp.23.558}. The goal of this architecture is to host applications in an edge cloud close to the base station and thus near to the clients in order to reduce end-to-end latency. The proposed architecture of this paper complements the \emph{3GPP} standard towards convergence of computing and connectivity across the whole cloud continuum.

\section{Holocom use case}
\label{sec:holocom}

\begin{figure}
    \centering
    \includegraphics[width=\linewidth]{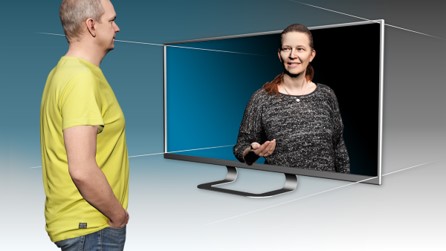}
    \caption{Holographic Communication}
    \label{fig:use_case}
\end{figure}

The \emph{6G NeXt} project is going to implement a real prototype of a future holographic communication service. The idea and goal of the use case called \gls{hc} is to research and develop a 3D video communication system based on true \gls{h3d} displays, as shown in Figure~\ref{fig:use_case}. We aim for realistic and natural remote communication between individuals enabled by \gls{h3d} technology invented by one of our partners. This technology creates light information in space in the same way as real objects would do, to allow natural viewing for the human eye including continuous depth, eye-focus, and convergence. This enables a natural audiovisual communication experience by reconstructing the call counterpart as a live and true 3D hologram in 3D space. Integrated eye tracking allows free movement and changing of perspective with little to no noticeable delay. Instead of using a \gls{h3d} display \gls{vr} headsets could be optionally used, although this limits options to capture the person's head and face for presentation on the other side.

The \gls{hc} use case requires 3D capturing of real persons in real-time, performed on each of the participating client systems. The 3D data will be transformed into a live 3D model to be synthesized into a virtual environment within a 3D engine running on the edge domain realized as a headless \emph{Unity} 3D rendering service~\cite{unity}. 

A \emph{Unity 3D specific plugin} to capture configurable 3D views compatible for \gls{h3d} is part of the development. The generated data is finally transferred to the other client using real-time streaming technologies such as WebRTC\cite{webrtc} or RTSP\cite{rtsp}. To enable this type of transfer, the captured \gls{h3d} compatible views are encoded into an image-based format. On the client side, the incoming 3D data is then computed into holograms and presented on the \gls{h3d} display or alternatively on a \gls{vr} headset.

The generated \gls{h3d} views depend on the real eye location of the observer looking at the \gls{h3d} display. Thus, with changing eye location, the \gls{h3d} views and accordingly the holograms need to be updated with high frequency. This underlines the challenging low delay and high bandw\textbf{}idth requirements that we expect in the \gls{hc} use case. Extremely low delay combined with high bandwidth enables a real-time eye movement to \gls{h3d} display photon latency enabling a realistic experience highlighting the requirement to 6G technology. Higher bandwidth clearly allows for faster transfer time to reduce delay and higher quality 3D data to be transferred to increase overall experience and quality of the person representation/reconstruction in full 3D.

Some key research will also be done in the field of 3D capturing with several cameras of different types. Normal RGB sensors to \emph{time-of-flight} or \emph{depth-from-stereo} based depth capturing methods will be considered in the research with the result of generating a live, high-resolution 3D model of a person. This will be based on the technology of one of our partners, which already enables high-quality 360 degree live scanning, producing more than 10 terabytes of data per minute. Key challenges will be clean capturing of, e.\,g., face details, hair, or glasses. Finally, audio capturing and hologram playout synchronization will be challenges to be solved.

\section{Concept of a Multi-level Split Computing Architecture}
\label{sec:concept}

In this section, we outline key concepts of our proposed \gls{hc} architecture.

\subsection{Remote Rendering}

\emph{Remote rendering} is an approach for migrating graphical computation tasks from consumer devices to remote servers, allowing even complex and computation-intensive 3D applications to be experienced on devices with limited processing capabilities, such as those of TVs. These devices typically have the capability to play high-resolution videos up to 4K or 8K, but are not designed for running applications with complex 3D rendering, which requires GPU acceleration.

In remote rendering, the viewport of the virtual camera within the 3D experience is rendered headless on the server and encoded as a video/audio stream, which is then delivered to the user device, which only needs to play the stream. User interactions, which can vary between different device classes, such as mouse/keyboard input on desktops, touch inputs on mobile devices, remote control on TVs, and motion control on \gls{ar}/\gls{vr} displays, must be captured and sent to the remote rendering server, which triggers the received events on the underlying rendering engine as if they were received from a connected input device.

The main challenge of remote rendering is reducing the end-to-end (E2E) latency while maintaining image quality. This latency, also known as \emph{Motion-to-Photon (MTP)} or \emph{Click-to-Photon (CTP)} latency, represents the delay until a user interaction is reflected on the display. Many factors affect the end-to-end latency and the quality of a remote rendering session under various conditions and constraints, which are summarized below:

\begin{itemize}
    \item Network: WiFi, Ethernet, LTE, 5G
    \item Browser: All modern browsers supporting WebRTC (Chrome, Firefox, Safari, Edge)
    \item Platform: Mobile, Desktop, TV, XR/VR, Holographic display
    \item Deployment: On-premise, Edge, Cloud or any combination of them
    \item Video codecs: VP8, VP9, AV1, H264, H265 
    \item Audio codecs: Opus, AAC
    \item Encoding/Decoding: Hardware Encoding/Decoding, Software Encoding/Decoding
    \item Server Resources (GPU/CPU/Memory): Various combination of accelerated compute instances provided by Hyperscalers
    \item Graphics Engine: Unity, Unreal Engine, NVIDIA Omniverse
    \item Protocol: WebRTC, HTTP3/QUIC
\end{itemize}

\begin{figure}[htbp]
    \centering
    \includegraphics[width=\linewidth]{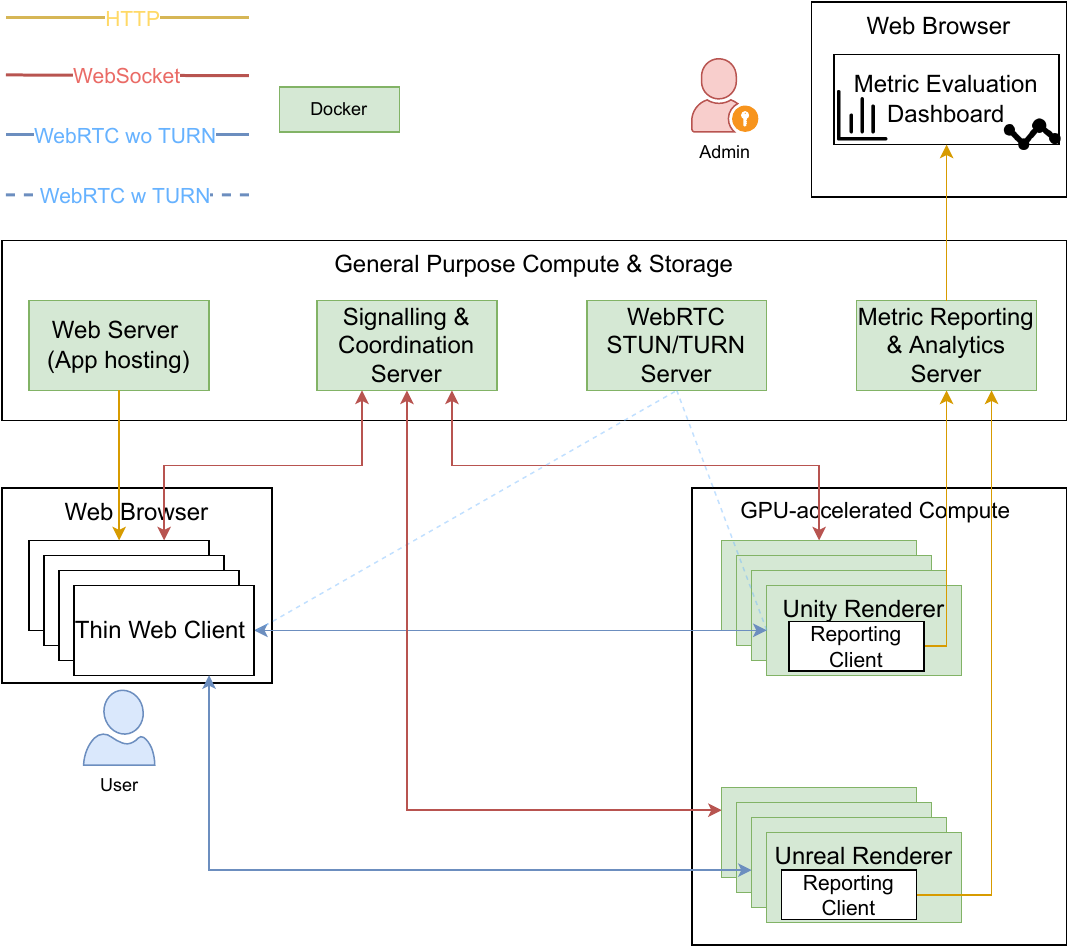}
    \caption{Remote Rendering Architecture}
    \label{remote-rendering-architecture}
\end{figure}

Figure~\ref{remote-rendering-architecture} shows the overall architecture of remote rendering. In order to reduce the end-to-end latency mentioned above, it is important to run the rendering instances close to the consumer device. This is facilitated by the edge computing capability of 5G networks and optimized in future 6G networks by supporting optimized and dynamic deployment of computation tasks across various cloud deployments.  

\subsection{High-Performance Communication and Computing Backbone}

The performance and efficiency of a new network generation will be determined by high-performance radio interfaces with application-optimized radio protocols as well as by ultra-fast software stacks, intelligent media processing, and the deep integration of \gls{ai} to optimize the overall system.
In the following, the key technologies for a joint communication and computing infrastructure are presented. 

\subsubsection{Split Computing}

In the \glspl{hc} use-case, capturing and rendering processes are distributed between a client device, edge, and central cloud deployments.
Distribution of computation tasks in the backbone layer and network based on \glspl{kpi} enhances the concept of edge computing and provides new possibilities for future application. The core of this concept is to match the requirements of an application such as \gls{hc} with the most suitable resources that are available based on \glspl{kpi}.

\subsubsection{Cross-Layer Metrics Function}
\label{cross_layer_metrics}

\begin{figure}
    \centering
    \includegraphics[width=.35\linewidth]{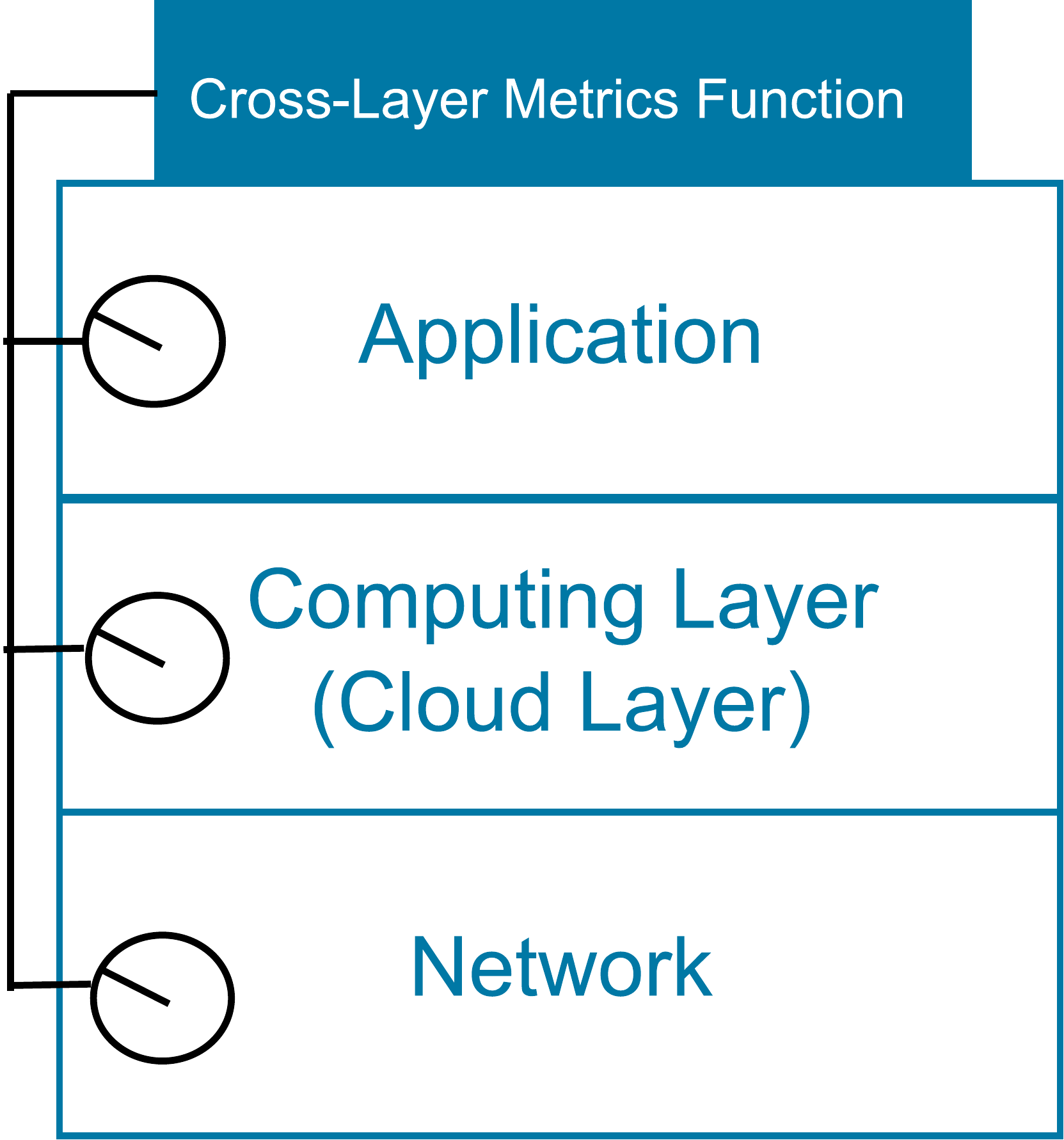}
    \caption{Cross-Layer Metrics Function}
    \label{fig:architecture2}
\end{figure}

Before a computing task such as rendering can be assigned to a certain computing infrastructure, it needs distinct knowledge about available resources and free capacities. A new proposed metric function aims at gathering metrics and measurement data from three different layers of the overall system, as shown in Figure~\ref{fig:architecture2}:

\paragraph{Application Layer} One part of the metric function collects data that are specific for \gls{hc} application. Typical data for the capturing part of the application include but are not limited to video texture resolution, mesh polygons number, mesh bit rate, and audio bit rate.
Rendering-specific data encompasses, e.\,g., the number of virtual cameras in the scene, used graphics engine,  rendering frame rate, or average video decoding time.
\paragraph{Computing Layer} This part of the metric function collects performance data of the computing infrastructure such as available CPU or GPU in the client device but also in the distributed edge-to-cloud environments.
\paragraph{Network Layer} The metric function also measures typical network characteristics such as throughput, latency, jitter, and packet loss between \gls{hc} clients and various instances of the rendering framework.

All data gathered by the cross-layer metric function are used for performance management of the \gls{hc} application but also for the entire computing and communication backbone and the optimization of the overall system.

\subsubsection{Discovery Service Function}
\label{discovery-function}

A \emph{Discovery Service}~\cite{baeldung} is a typical part of a microservice architecture. Services such as the \gls{hc} rendering framework might change a number of instances and locations in a virtual and containerized environment over time. The \emph{Discovery Service} provides \gls{hc} clients with the network address of a remote rendering service. Therefore, every instance of a remote rendering service has to register itself with the \emph{Discovery Service} by sending its current address. This is typically the dynamically assigned network path.

The proposed architecture takes the concept of \emph{Discovery} one step further and enables a client to send parameterized \emph{Discovery Request}. A client might request an instance of a rendering service that fulfills certain \glspl{kpi}. These \glspl{kpi} might be regarded to bandwidth or latency requirements but also to \gls{qos} classes or service costs. The proposed discovery service function keeps track of the instance addresses but takes also data of the \emph{Cross-layer Metric Function} into account in order to provide \gls{hc} clients with the most suitable rendering framework instance in order to assure an optimal service experience.   

\subsubsection{Service Broker Function} \label{service-broker-function}
A service broker helps services to find, e.\,g., free cloud capacities and also to start a new instance based on requirements described with \glspl{kpi}. It is an enhanced version of a discovery service. When the remote rendering framework recognizes a decreasing service quality due to missing memory and CPU/GPU in the current cloud environment it might ask the Service Broker Function for a new deployment possibility. The Service Broker Function performs two tasks: 
First, it recommends a suitable cloud resource with sufficient connectivity for increasing service capacities.
Second, it helps the service to deploy and start a new service instance using Docker~\cite{docker} and automate infrastructure tools. Also, the network connectivity might be configured using slicing capabilities or network APIs, e.\,g., those specified in CAMARA~\cite{camara}.

\subsubsection{Comprehensive Architecture View} \label{overal-view}
\begin{figure}
    \centering
    \includegraphics[width=\linewidth]{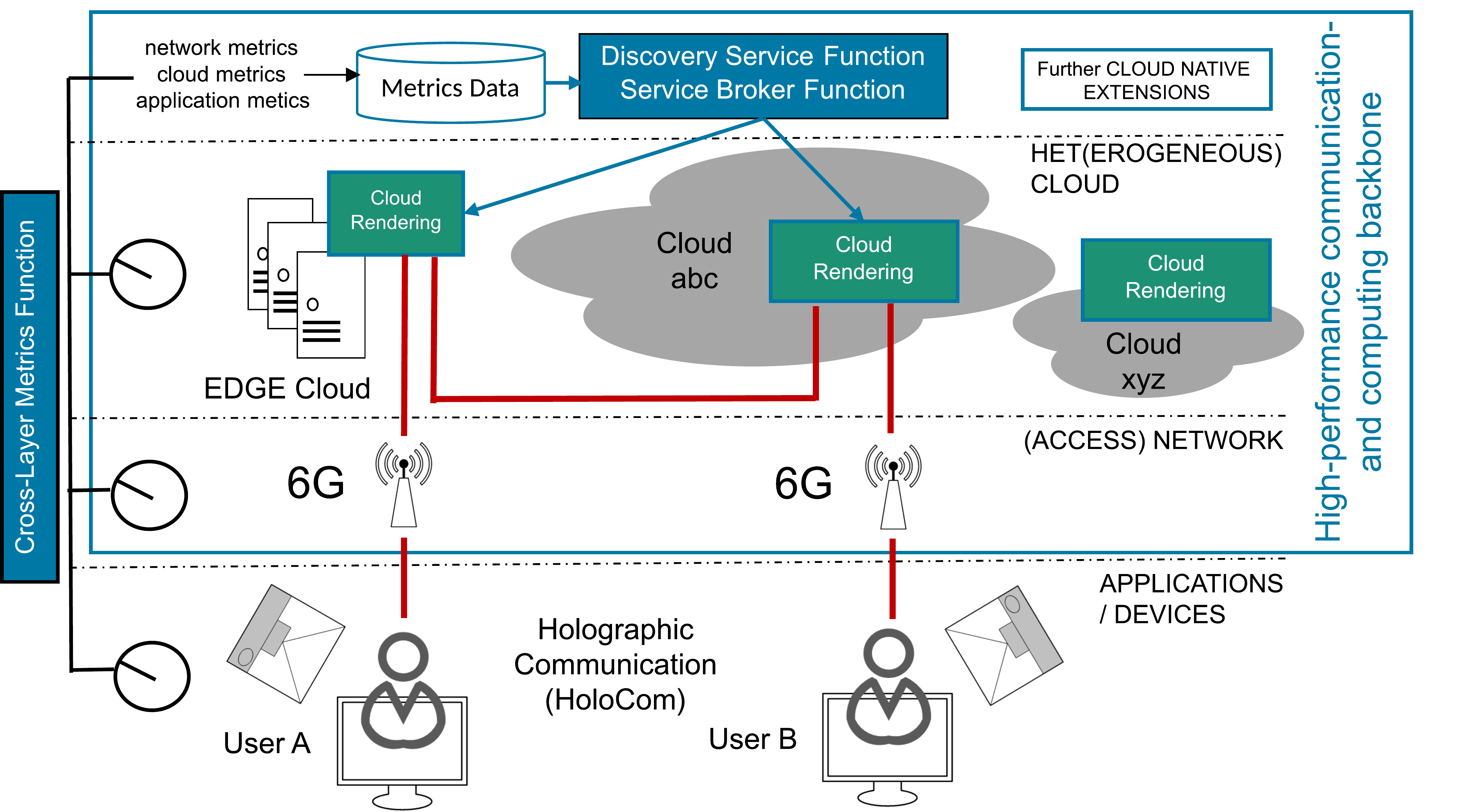}
    \caption{High-performance communication-and-computing backbone for holographic communication}
    \label{fig:architecture}
\end{figure}

The \gls{hc} use-case running on the proposed architecture is shown in Figure~\ref{fig:architecture}: 

The \gls{hc} user clients contact the \emph{Discovery Service Function} providing their minimal requirements regarding latency, possibly caused by the network and remote rendering service. The \emph{Discovery Service Function} provides user client \emph{\textsf{A}} with the address of a remote rendering service located in the edge-cloud while client \emph{\textsf{B}} is connected to a remote rendering service in central cloud deployment \emph{\textsf{abc}}. 
The \emph{Discovery Service Function} provides its recommendation based on metrics data gathered by the \emph{Cross-Layer Metrics Function}.
The remote rendering service starts a new instance in cloud \emph{\textsf{xyz}} with the help of the \emph{Service Broker Function}.
The architecture is characterized by three new functions added, the \emph{Cross-Layer Metrics Function}, the \emph{Discovery Service Function}, and the \emph{Service Broker Function} as extensions to standard cloud environments. Further extensions are going to be described in the next section.

The classic separation of network and computing infrastructure in layers is eliminated and replaced by a modular approach. The clear distinction between computing infrastructures also starts to blur. The \emph{High-performance communication-and-computing backbone} enables a flexible split of computing tasks according to the requirements of the application. And applications get all information and support to decide where to run, where to connect or to be deployed to.  

\subsection{Edge-to-Cloud Serverless Software Extensions}

The proposed system leverages existing serverless software platform abstractions for compute, data, and message management in the 6G edge-to-cloud continuum to handle irregular and supplementary tasks such as metadata management.

\subsubsection{Compute}
In addition to the resource-intensive tasks of cloud holographic rendering, the metrics functions that provide ancillary system measurement data must be deployed in a scalable, isolated, and resource-efficient manner throughout the edge-to-cloud continuum.
Specifically, metrics must be collected at the client side, within the network, and at the remote rendering location.
To deploy this irregular task with both geographical scalability and isolation between tenants, we rely on the \gls{faas} compute paradigm, where small, stateless functions can be deployed in an event-driven manner.
Using the \emph{tinyFaaS}~\cite{paper_pfandzelter_tinyfaas} lightweight edge \gls{faas} platform, we can efficiently provide a software infrastructure to collect critical performance metrics data.
Additionally, aggregation and filtering of metrics data can be performed as \gls{faas} functions in the edge-to-cloud continuum to reduce the amount of data that must be transmitted to the service broker.

\subsubsection{Messaging}
After collecting metrics data in geographically distributed locations, it must be distributed efficiently to where it is needed for decision-making.
To address this challenge, we leverage the \gls{pubsub} paradigm where services and application components can subscribe to relevant metrics data that individual metrics functions publish.
Crucially, through the introduction of geographical awareness and intelligent message routing as well as the early aggregation and filtering of metrics data, we reduce the overall network strain of distributing metrics data.
For these tasks, we rely on the \emph{GeoBroker}~\cite{paper_hasenburg_geobroker,paper_hasenburg_disgb} \gls{pubsub} middleware.

\subsubsection{Data}
While the holographic data that is streamed through the system may be ephemeral, there is supporting data and metadata that is crucial to the operation of the system.
This data includes client authentication information, as well as other critical data that must be accessed in a low-latency manner.
Furthermore, clients and services may move fluidly throughout the edge cloud and the real world, and data copies must move with them.
In order to replicate such metadata at the client side and the rendering service, and in order to do so even with service or client movement, we rely on the abstractions provided by the \emph{FReD}~\cite{pfandzelter2023fred,hasenburg2019fbase,hasenburg2020towards} fog data management system.
Using application-controlled data replica placement, we can ensure low-latency metadata access without impairing user experience, adhering to \gls{qos} and \gls{qoe} objectives.

\subsection{QoE Alignment}
The user experience is a central factor for the subsequent satisfaction of users in immersive communication. In the context of \gls{hc}, humans may be sensitive to errors in the displayed information or more subtle impairments due to delay. Along with this, higher values of \gls{qos} do not necessarily ensure higher \gls{qoe}~\cite{agiwal2016next}. Hence, it is necessary to conduct detailed user studies on the perceived quality of \gls{hc}. The perceived quality of immersive communication is to be evaluated through subjective as well as objective measures. The subjective measures will be assessed using questionnaires related to constructs such as co-presence, immersion, and realism, as well as in terms of audiovisual quality, which is often measured with a 5-point Absolute Category Rating resulting in a mean opinion score (MOS)~\cite{aseeri2021influence}. Along with this, we will record conversational and behavioral data of the participants for indirect evaluation of user experience~\cite{skowronek2022quality}. The \gls{qoe} of users is related with underlying characteristics of the end device and the network and edge technology used (\gls{qos}) to identify respective \glspl{kpi}.

A joint architecture for communication and computation is proposed for \gls{hc} systems, where the computation of the data from capturing to rendering is distributed across cloud deployments and networks. The parameters of transmission (e.\,g., codecs, compression ratio and bandwidth, delay, and possible data loss) will be collected to develop a prediction model, relating the aforementioned \gls{qoe} indicators with such service-related KPIs. Existing state-of-the-art monitoring models such as the framework of the ITU-T P.1203 and P.1204 model series will be used as a conceptual basis, adapted to our context~\cite{robitza2018http, raake2020multi}. Such \gls{qoe} models can serve to automatically measure the current \gls{qoe} of the application and inform about root-causes, enabling AI-assisted communication system to allocate adequate resources for maintaining user satisfaction of the \gls{hc} systems.

At the receiver side, the rendering of the communication partner is distributed across the computing resources in the cloud/edge networks for efficient processing of the data. However, synchronization of the processed data stream from distributed media processing functions is important for rendering a communication partner. If the data streams from the distributed media processing functions are not synchronous, it might lead to misaligned representations of the rendered object~\cite{clemm2020toward}. Also, it is highly desirable to have a synchronous audio signal along with the visual representation to support better \gls{qoe} in the \gls{hc} systems. Along with this, undesirable values of transmission parameters (e.\,g., higher data loss, longer delay) severely impact the quality of rendered objects and produce possible artifacts such as blurring, freezing, etc., which negatively impact the user experience.

\subsection{AI-Supported Flexible Mobile Communication Platform}
Since wireless networks are now rapidly growing with massive network items, wide bandwidth, and high data rates, it is necessary to take the uncertainty and the changing environment of the radio access network into consideration to improve the system reliability, \gls{qos} and even \gls{qoe}. Recently, \gls{ml} models are increasingly being used to solve those non-linear problems, e.g., network slicing, channel prediction, or \gls{qos} optimization. Because of those advantages for adapting the spatiotemporal dynamics in advance based on the pre-trained model on a huge amount of historical channel data~\cite{QZ_10012940,QZ_10012951,QZ_8813020,QZ_9354041,QZ_9569310,QZ_9044427,QZ_9860403}. In communication networks with moving objects, the experienced \gls{qos} is affected by various factors, such as \gls{ue} density, interference, mobility, handover, and roaming transitions. The prediction of worse network services at certain locations due to one or many of those factors is one of the key requirements for critical applications with high demand of \gls{qos}/\gls{qoe}. A recent approach has combined the most efficient \gls{ml} methods, that an \gls{lstm} and a \gls{drl} are used to conduct online decentralized testing at the UE pairings, which gives resource units the ability to allocate bandwidth and make decisions about packet scheduling \cite{QZ_9146836}. Further, \gls{fdrl} is proposed for a UAV-aided vehicular network, which is examined to improve connection and minimize latency \cite{QZ_alsulami2022federated}. \gls{ml} methods are used at every level in distributed federated systems, which reduces latency and improves accuracy, which are the most important indexes of the quality of \gls{hc}. 

\begin{figure}
    \centering
    \includegraphics[width=1\linewidth]{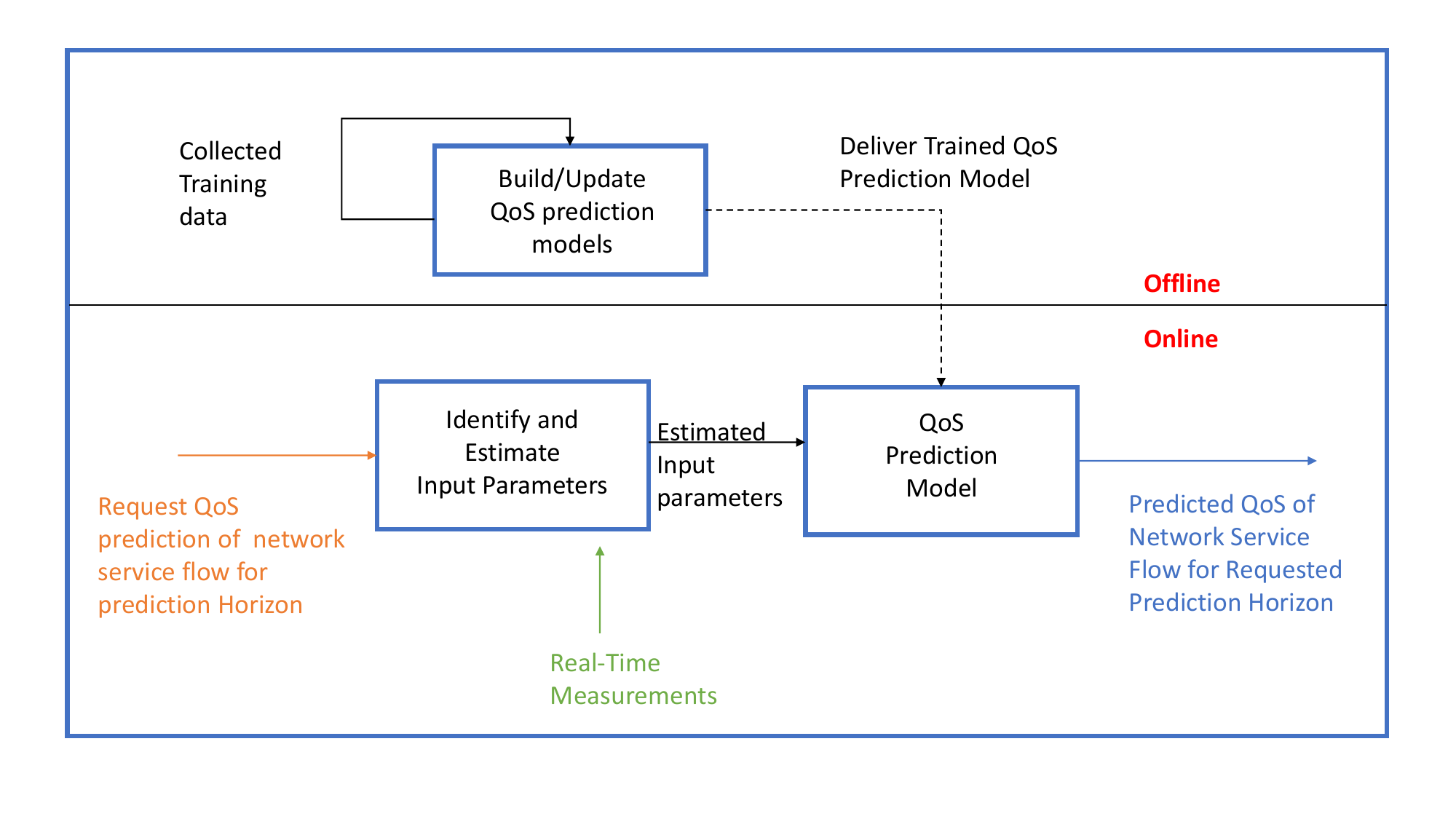}
    \caption{Overview of \gls{qos} prediction scheme for network service flow}
    \label{fig:mlcsi}
\end{figure}

An overview of such a \gls{ml}-based prediction model is shown in Figure \ref{fig:mlcsi}. 
    \subsubsection{Offline phase}
    In this phase, \gls{qos} data are collected using network and application monitoring tools, which are processed to train and periodically update a prediction model. Different data and input features need to be used for an accurate prediction of different \gls{qos} parameters. The objective of this phase is to leverage machine learning techniques to establish connections between input features and their corresponding predicted parameter and estimate the anticipated value of a \gls{qos} parameter with greater accuracy. This phase is conducted offline and does not recur for each individual prediction request, although it may be periodically re-executed based on the latest collected data. It may be necessary to develop distinct prediction models for each \gls{qos} parameter and/or for Uplink and Downlink signals. The trained models are then made available to the online engine.
    \subsubsection{Online phase}
    This phase is triggered by each prediction request. Based on the type of request, the relevant input features are identified, and their expected values are projected for the prediction horizon specified by the application. Real-time measurements are also gathered by the appropriate network entities to determine the present and future values of the input features for the given service flow. In the second stage of this phase, the anticipated values of the input features are inputted into the trained \gls{qos} prediction model, which calculates the anticipated \gls{qos} of a service flow for the requested prediction horizon.


In \Cref{fig:architecture}, a user connects to the \emph{remote rendering} service instance by means of a wireless link. However, on the application layer, the communication link may be divided into different traffic types, such as video, audio, control, etc. Every of these traffic types has its dedicated requirements on throughput, reliability, latency, and others. The variety of the traffic types becomes even larger if some other cloud-based applications are active in parallel to a \gls{hc} call. The very heterogeneous requirements of different applications and in particular different traffic types inside a single application require high flexibility of the underlying wireless network in order to provide optimal \gls{qoe} to the user.

Especially, interactive applications such as \gls{hc} require high link reliability in conjunction with low latency. This means, that the approach of retransmission of lost packets is not appropriate, but the packets should arrive on time with the given loss probability. Besides, the well-established adaptation of \gls{mcs} to the quality of the wireless channel does not provide the required flexibility, since the jump between two \glspl{mcs} is too high in terms of throughput change.

To overcome this shortage, we propose the utilization of fountain codes, which belong to the family of rateless codes. Here, an arbitrary number of sub-codewords is generated out of a payload packet, but only a certain number of those is required to decode the packet. Thus, the coding rate can be flexibly adjusted.
%
%
However, these codes do not include \gls{fec} capability thus are not able to detect nor to correct corrupted sub-codewords. In order to overcome this shortage, a two-stage coding approach can be applied. In the first stage, \gls{fec} codes are utilized to turn the wireless channel into a packet erasure channel. In the second stage, fountain coding is utilized to correct occurred packet drops. Authors in \cite{sm_karrenbauer2020study} showed, that the joint erasure and error-correcting scheme shows better performance as compared to the one-stage approach.

The main advantage of fountain codes is their flexibility. Based on predicted channel quality, the coding rate can be adjusted ``on the fly'' in order to meet the required reliability \cite{sm_melnyk2022wireless}. Since the coding rate is not related to \gls{mcs} settings, it can be re-adjusted for any piece of payload. It provides the possibility to divide the heterogeneous traffic types into separate communication queues. After, based on predicted channel quality and the requirements of each queue, the coding scheme can be flexibly adjusted to meet the \gls{qos} requirements on the one hand, and to provide optimal utilization of communication channels on the other hand.


\section{Conclusion}
\label{sec:conclusion}

Sixth generation networks should bring mobile communication to the next level making the connectivity omnipresent for commercial and private usage. In order to realize this vision, new technological ideas are required. In this paper, we presented an approach and an architecture of joint communication and computing infrastructure. The proposed high-performance backbone layer in combination with additional cloud native services allows the distribution of processing power in all levels inside the communication network, which reduces the demands on the client devices on the one hand, and simplifies the realization of collaborative applications on the other hand. 
With \gls{hc} a demanding use case was chosen to provide real-world requirements on latency, throughput, reliability, availability of computation power etc. The goal of the project \emph{6G NeXt} is to go beyond the conceptual evaluation of proposed technologies towards a functional proof of concept based on the proposed \gls{hc} use case.

\section*{Acknowledgements}


Funded by the Bundesministerium für Bildung und Forschung (BMBF, German Federal Ministry of Education and Research) -- 16KISK174K, 16KISK176, 16KISK177, 16KISK178, 16KISK179, 16KISK180, 16KISK182, 16KISK183.

\balance

\printbibliography



\end{document}